\documentstyle[12pt]{article}
\newcommand{\be}{\begin{equation}}
\newcommand{\en}{\end{equation}}
\newcommand{\bea}{\begin{eqnarray}}
\newcommand{\ena}{\end{eqnarray}}

\newcommand{\hbo}{\hbox to 1 true cm {\hfill } }

\newcommand{\Tr}{\hbox{Tr}}

\input psfig
\def\dslash{\partial\kern-.5em\slash}
\def\kslash{k\kern-.5em\slash}

\begin{document}
\baselineskip=16truept
\vglue 1truecm

\vbox{
\hfill UNIT\"U-THEP-2/1993, March 9, 1993
}

\vfil
\centerline{\bf The non-trivial phase of $\phi^{4}$-theory in a
renormalisation }
\centerline{\bf group invariant approach }

\bigskip
\centerline{ K.\ Langfeld, L.\ v.\ Smekal, H.\ Reinhardt }
\medskip
\centerline{Institut f\"ur theoretische Physik, Universit\"at
   T\"ubingen}
\centerline{D--7400 T\"ubingen, Germany }
\bigskip

\vfil
\begin{abstract}

Massless $\phi^{4}$-theory is investigated in zero and four
space-time dimensions.
Path-integral linearisation of the $\phi ^{4}$-interaction defines
an effective theory, which is investigated in a loop-expansion
around the mean field. In zero dimensions this expansion converges
rapidly to the exact potential obtained numerically.
In four dimensions its lowest order (mean-field approximation)
produces a real and convex effective
potential. Two phases are found. In one the renormalisation group
improved one-loop effective potential is recovered as the leading
contribution near the classical minimum. This phase, however, is unstable.
The second (precarious) phase is found to have lower vacuum energy density.
In this phase a dynamical mass is generated.
The results are renormalisation group invariant.

\end{abstract}

\vfil
\hrule width 5truecm
\vskip .2truecm
\begin{quote}
$^1$ Supported by DFG under contract Re $856/1 \, - \, 1$
\end{quote}
\eject
%\tableofcontents
%
%
$\phi ^{4}$-theories enjoy a broad span of applications ranging
from solid state physics~\cite{stn72} to cosmological
models~\cite{ko90}. Of particular importance is the role played by
the vacuum structure of $\phi ^{4}$-theory in the standard model
of electro-weak interactions. In this context it is generally believed
that the vacuum of $\phi^{4}$-theory spontaneously breaks the weak isospin
symmetry (see e.\ g.\ \cite{ch84}), thus giving rise to a number of
significant phenomenological consequences - for example it generates,
via the Higgs-Mechanism, the masses of the electro-weak gauge bosons.

The effective potential of a relativistic quantum field theory is
a convenient tool for studying the vacuum structure of a theory.
In particular, its global minimum value is the vacuum energy
density~\cite{co73}. Another well-known property of the effective potential
of $\phi^{4}$-theories is that it is a real and convex function of
the fields~\cite{ilo75}.

In four space-time dimensions renormalisation
induces an arbitrary energy scale (renormalisation point).
An important feature of Quantum-Field-Theory is renormalisation group
invariance which ensures that physical observables do not depend on this
scale.

Unfortunately, calculation of the effective potential requires
a knowledge of the full generating functional for Green's functions,
a quantity which usually cannot be calculated exactly.
Even for $\phi ^{4}$-theory one has to resort to either
numerical simulations~\cite{cal83} or approximation schemes. One such
scheme is the conventional loop expansion, which is also an expansion
in the coupling constant~\cite{co73}. However,
it is well known that the loop expansion in massless $\phi ^{4}$-theory
breaks down near the minimum of the potential, where it yields
a potential which is neither real nor convex. This disease of the one-loop
effective potential can be cured however, by applying renormalisation
group arguments and the renormalisation group improved effective potential
thus obtained is convex for small
scalar fields, but ceases to exist for large fields~\cite{co73,ka92}.
Perhaps even worse are the indications that this phase of
$\phi ^{4}$-theory is trivial (i.\ e.\ at best it becomes free)
due to renormalisation effects~\cite{aiz81}.
This is particularly unpleasant because
many phenomenologically successful models are based on a non-trivial vacuum
structure of $\phi^{4}$-theory.

In the {\it massive \/} case it was shown by Stevenson that a variational
estimate of the effective potential exhibits a non-trivial
"precarious" phase~\cite{ste84}, so called because the bare coupling is
negative and infinitesimally small. Stevenson argued that this phase
escapes the triviality bounds from lattice calculations. However,
it seems likely that
the effective potential obtained in this variational approach is not
renormalisation group invariant
\footnote{This will be demonstrated in more detail elsewhere~\cite{la93}. }.

In this letter we use a different approach to study especially
{\it massless \/}
$\phi ^{4}$-theory, based on a path-integral linearisation of the
$\phi ^{4}$-interaction. We evaluate the effective potential in a
modified loop expansion around the mean field. This expansion
converges rapidly in zero dimensions where it is compared
to the exact numerical result. In four space-time dimensions
we construct the effective potential in leading order (mean-field
approximation) using an appropriate renormalisation prescription.
We find two phases. In one the renormalisation group improved
one-loop effective potential is recovered without resorting to a small
coupling expansion. This is the phase which is believed
to be trivial. In addition to this phase we find a non-trivial one
with lower energy density which must therefore form the true vacuum.
In this phase a mass for the scalar particle is dynamically generated.
The results for the vacuum energy density and the mass of the
scalar field are shown to be renormalisation group invariant.

The Euclidean generating functional for Green's functions of the
$\phi ^{4}$-theory is given in $D$ space time dimensions by
\be
Z[j] \; = \; \int {\cal D} \phi \; \exp \{ - \int d^{D}x \;
\bigl( \frac{1}{2} \partial _{\mu } \phi \partial _{\mu } \phi \, + \,
\frac{ m^{2} }{2} \phi ^{2} \, + \, \frac{ \lambda }{24} \phi ^{4}
\, - \, j (x) \phi (x) \, \bigr)  \} \; ,
\label{eq:1}
\en
where $m$ is the bare mass of the scalar field and $\lambda $ the
bare coupling constant.
$j(x)$ is an external source for $\phi (x)$ which provides access to the
Green's functions. The effective action
is defined as the Legendre transform
\be
\Gamma [ \phi _{c}] := - \ln Z[j] + \int d^{4}x \;
\phi _{c}(x) j(x) \; , \hbo
\phi _{c}(x) :=  \frac{ \delta \, \ln Z[j] }{ \delta j(x) } \; .
\label{eq:2}
\en
The effective potential $U(\phi _{c}) = \frac{1}{V} \Gamma [\phi_{c}
=\hbox{const.} ] $ is obtained from (\ref{eq:2}) for a
space-time independent classical field $\phi_{c}$.

In order to calculate $Z[j]$ we linearise the $\phi ^{4}$-interaction
by means of an auxiliary field $\chi (x)$
\be
Z[j] \; = \; \int {\cal D} \phi \; {\cal D} \chi \;
\exp \{ - \int d^{D}x \; [ \frac{6}{\lambda } \chi^{2} (x)
\, + \, [\frac{ m^{2} }{2} - i \chi (x) ] \phi ^{2}(x)
\, - \, j(x) \phi (x) ] \} \; ,
\label{eq:3a}
\en
where constant factors have been ignored.
The integral over the fundamental field $\phi $ is then
easily performed, yielding
\bea
Z[j] &=& \int {\cal D} \chi \; \exp \{ - S[\chi,j] \} \; ,
\label{eq:3} \\
S[\chi,j] &=& \; \frac{ 6 }{ \lambda } \int d^{4}x \; \chi^{2}  \, + \,
\frac{1}{2} \Tr _{(R)} \ln \, {\cal D}^{-1}[\chi ] \, - \, \frac{1}{2}
\int d^{4}x \; d^{4}y \; j_{x} {\cal D}[\chi ]_{xy} j_{y} \; ,
\label{eq:5} \\
{\cal D}^{-1}[\chi ]_{xy} &=&
(- \partial ^{2} + m^{2} - 2i \chi (x) ) \delta _{xy}  \; .
\label{eq:6}
\ena
The trace $\Tr _{(R)}$ extends over space-time, and the subscript
indicates that a regularisation prescription is required.
For simplicity we use cutoff regularisation, but any other method
(e.\ g., dimensional regularisation) would also work.
For this $\chi $-theory the classical field $\phi _{c}(x)$
(\ref{eq:2}) becomes
\be
\phi_{c}(x) \; = \; \int d^{4}y \; \langle {\cal D}[\chi ]_{xy}
\rangle \; j_{y} \; ,
\label{eq:6a}
\en
where $\langle \ldots \rangle $ denotes averaging over the auxiliary
field $\chi $ with the weight factor $\exp ( -S[\chi , j ] ) / Z[j] $.
Evaluation of the $\chi $-integral by the method of stationary \break
phases~\cite{be78} leads to the mean-field approximation
\be
\frac{ \delta S[ \chi , j] }{ \delta \chi (x) }
\vert _{\chi = \chi _{0} } \; = \; 0 \; .
\label{eq:m1}
\en
Expanding the action around this mean-field solution $\chi _{0}(j)$
provides a modified loop expansion. With $\chi ' = \chi - \chi_{0}$,
this is
\bea
Z[j] &=& \; \exp \{ - S[\chi _{0}(j),j] \} \;
\exp \{ - \sum _{\nu \ge 3} \frac{1}{2}
S^{(\nu )}[ \frac{ \delta  }{ \delta J },j ]  \}
\label{eq:4} \\
& \phantom{=} &
\int {\cal D} \chi' \; \exp \{ - S^{(2)}[\chi',j]
\, + \, \int d^{D}x \; J \chi' \} \; \vert _{J=0} \; ,
\nonumber
\ena
where
\[
S^{(\nu )} [\chi',j ] \; = \; - \frac{1}{ \nu } \; \Tr \{
{\cal D} [\chi _{0}] \, (2i \chi') \}^{\nu }  \; - \;
\int d^{4}x \; d^{4}y \;
j_{x} \{ \, [ {\cal D} [\chi _{0}] (2i \chi')] ^{\nu } {\cal D}[\chi _{0} ]
\, \} _{xy} j_{y}
\]
is the $\nu $-th order $\phi $-loop with non-vanishing external source
$j_x$.
In particular $S[\chi ]$ of the $\phi ^{4}$-theory
is minimal for a purely imaginary $\chi _{0}$. We therefore deform the
path of integration in (\ref{eq:3}) within the regime of analyticity of
$S(\chi )$  to include $\chi _{0}$ before performing the expansion which
leads to (\ref{eq:4}).
The integration contour and the saddle point are uniquely determined
by this procedure, which is known as the stationary phase
approximation with complex saddle points~\cite{be78}.

We have tested this modified loop expansion in zero dimensions $(D=0)$
where $\int {\cal D} \chi $
becomes an ordinary integral and the kinetic term $\partial ^{2}$
is absent from the propagator ${\cal D}$. In this case the effective
potential (which is simply the effective action here) can be
readily evaluated numerically. The important observation is that the
modified loop-expansion (\ref{eq:4})
provides a rapidly converging expansion for $Z[j]$.

In figure 1 we compare the exact effective potential
$U(\phi _{c})$ with the leading order (mean field approximation)
and next to leading order in the modified loop expansion
for zero mass $(m=0)$. It is seen that this modified expansion
converges rapidly and that the mean-field approximation provides
an adequate description of $U(\phi _{c})$ particularly near the minimum,
the region of physical interest.
We have checked that the potential is real and convex even for the
case of an imaginary mass $(m^{2} < 0)$. This is quite an improvement
compared to the perturbative approach since in the standard loop expansion
spurious phase transitions occur (for small $\vert m^{2} \vert $) implying a
break down of the perturbative approach for small fields~\cite{la93}.
For large masses
reasonable results are achieved by the conventional loop expansion.
Similarly, if $(- m^{2})$ is large, an improved scheme can be successfully
applied~\cite{be83}.

In four space-time dimensions renormalisation makes the above
investigations more complicated.
For a constant source $j$ (and thus for constant classical field
$\phi _{c}$) the action $S[\chi ,j]$ (\ref{eq:5})
is minimised by constant complex fields
$\chi _{0}=:\frac{i}{2}(M-m^{2})$, where $M$ is determined by solving
the equation of motion (\ref{eq:m1}).
For this first investigation we stick to the classical
approximation. At this level the generating functional is
$Z[j]=\exp \{ - S[\chi _{0}] \}$. To this order the classical field
is easily obtained from (\ref{eq:6a}), i.\ e.,
$\phi _{_c} (x) = \int d^{4}y \; {\cal D}[\chi _{0}] _{xy}\, j_y$.

Eliminating the source $j(x)$ in favour of the classical field $\phi _{c}$
the equation of motion (\ref{eq:m1}) can be cast into the form
\be
\frac{6}{\lambda } M \; - \; \frac{ 6 m^{2} }{ \lambda }
\; - \phi_{c}^{2} \; = \;
\int _{(R)} \frac{ d^{4}p }{ (2\pi )^{4} }
\frac{1}{ p^{2} + M } \; .
\label{eq:7}
\en
{}From the definition (\ref{eq:2}) the effective potential is
\be
U (\phi _{c}) \; = \; \frac{3}{2 \lambda } M^{2} +
\frac{1}{2 } \int_{(R)} \frac{ d^{4}p }{ (2\pi )^{4} } \;
\ln ( p^{2} + M ) \, - \, \frac{1}{2} M
\int_{(R)} \frac{ d^{4}p }{ (2\pi )^{4} } \frac{1}{ p^{2} + M } \; ,
\label{eq:8}
\en
where $M$ is a function of the classical field by (\ref{eq:7}).
Using the cutoff regularisation, the trace
$\Tr_{(R)}$ and the loop integral in (\ref{eq:8}) become
\bea
\int_{(R)} \frac{ d^{4}p }{ (2\pi )^{4} }
\; \ln ( p^{2} + M ) &=& \frac{1}{16 \pi^{2}}
\{ M \Lambda ^{2} + \frac{ M^{2} }{2} ( \ln \frac{ M }{ \Lambda ^{2} }
- \frac{1}{2} ) \} + O( \frac{1}{ \Lambda ^{2} } )
\; , \label{eq:9} \\
\int_{(R)} \frac{ d^{4}p }{ (2\pi )^{4} }
\frac{1}{ p^{2} +M } &=& \frac{1}{16 \pi^{2}}
\{ \Lambda ^{2} + M \ln \frac{M}{\Lambda ^{2} } \} \; ,
\label{eq:10}
\ena
where $\Lambda $ is the momentum cutoff. Renormalisation
is performed by absorbing the divergences of (\ref{eq:7}, \ref{eq:8})
into the bare mass and the bare coupling constant. Defining the
renormalised quantities $\lambda _{R}, \, m_{R}$ by
\vfill \eject
\be
\frac{6}{\lambda } \; + \; \frac{1}{16 \pi ^{2} }
\ln \frac{ \Lambda ^{2} }{ \mu ^{2} } \; = \;
\frac{6}{ \lambda _{R}(\mu ) }
\; , \label{eq:11}
\en
\be
\frac{6 m^{2} }{ \lambda } \; + \; \frac{1}{16 \pi ^{2} } \Lambda ^{2} \; = \;
\frac{ 6 m_{R}^{2}(\mu ) }{ \lambda _{R}(\mu ) }
\label{eq:12}
\en
finite results for (\ref{eq:7}), (\ref{eq:8}) are produced.
This is our crucial observation. Note that
the bare coupling $\lambda $ is negative and becomes infinitesimally
small, if the regulator $\Lambda $ goes to infinity. The scale
$\mu $ is an arbitrary subtraction point - later on we must check
that physical quantities do not depend on it.
The specific choice of the right hand side of (\ref{eq:11},\ref{eq:12})
is arbitrary. The form we have selected corresponds to certain
renormalisation conditions, which reproduce the
standard definitions of renormalised mass and coupling strength~\cite{co73}.
In terms of the renormalised quantities the equation of motion (\ref{eq:7})
and the effective potential (\ref{eq:8}) become
\be
\frac{6 M}{ \lambda _{R} } \; - \; \frac{6 m_{R}  }{ \lambda _{R} }
\; - \; \phi_{c}^{2} \; = \; \frac{ M }{ 16 \pi ^{2} }
\, \ln \frac{ M }{ \mu ^{2} }
\; , \label{eq:13}
\en
\be
U (\phi _{c} ) \; = \; \frac{3}{2 \lambda _{R} } M^{2} \; - \;
\frac{ M^{2} }{ 4 (16 \pi ^{2} ) } \, [ \ln \frac{ M }{ \mu ^{2} }
+ \frac{1}{2} ] \; .
\label{eq:14}
\en
This is the desired result since both equations (\ref{eq:13}) and
(\ref{eq:14}) are finite.
% We have also obtained here a renormalisation group invariant
% effective potential $U $.
It is in fact easy to verify that neither
$M$ (\ref{eq:13}) nor effective potential (\ref{eq:14}) depend
on the renormalisation point $\mu $.
This is because from (\ref{eq:11}, \ref{eq:12}), we have
\be
\beta (\lambda _{R}) := \mu \frac{ d \lambda _{R} }{ d \mu }
= \frac{1}{3\cdot 16 \pi ^{2} } \lambda _{R}^{2} \; , \hbo
\frac{ d }{ d \mu } \frac{ m_{R} }{ \lambda _{R} } \; = \; 0
\; . \label{eq:15}
\en
So we see that $U(\phi _{c})$ is finite and renormalisation group
invariant.
Note that the scaling coefficient $\beta _{0} =
\frac{1}{3 (4\pi )^{2} }$ in (\ref{eq:15}) is smaller than
the usual one obtained from perturbation theory.
However, the approach here is intrinsically non-perturbative, there is no
reason to expect $\beta (\lambda _{R})$ to agree with the conventional
power series expansion for small $\lambda _{R}$.

In order to make contact with the conventional
one-loop effective potential~\cite{co73}
we solve (\ref{eq:13},\ref{eq:14}) in the massless case $(m_{R}=0)$.
To this end we cast (\ref{eq:13}, \ref{eq:14}) in the form
\bea
M &=& \; \frac{ \lambda_{R} \phi _{c}^{2}/6 }{ 1 -
\frac{ \beta_{0} \lambda _{R} }{2} \ln \frac{ M }{ \mu ^{2} } } \; ,
\nonumber \\
U (\phi _{c}) &=& \frac{1}{4} \phi _{c}^{2} M \;
- \frac{ 1 }{ 8 (4\pi )^2 } M^{2} \; .
\nonumber \\
\ena
Both terms in $U (\phi_c )$ are separately RG invariant and near
the classical minimum $(\phi_c = 0)$ the first term dominates. So, in
this regime, the leading invariant contribution to $U $ is given by
\be
U (\phi _{c}) \; = \; \frac{ \lambda _{R} \phi _{c}^{4}/24 }{
1 - \frac{ \beta _{0} \lambda _{R} }{2}
\ln \frac{ \phi _{c}^{2} }{ \mu ^{2} } } \; .
\label{eq:16}
\en
This is precisely the renormalisation group improved one loop effective
potential given in~\cite{co73,ka92}.
Note that this standard result is obtained with the {\it
precarious \/}~\cite{ste84} renormalisation scheme given by equations
(\ref{eq:11},\ref{eq:12}), i.\ e.,
\be
\lambda \; = \; \frac{ \lambda _{R}(\mu ) }{ 1 \, - \,
\frac{ \beta _{0} \lambda _{R} }{2}
\ln ( \Lambda ^{2} / \mu ^{2} ) } \; \rightarrow \; 0^{-} \; .
\label{eq:m3}
\en
This behaviour is in fact hidden in the standard perturbative
approach which corresponds to an expansion of (\ref{eq:m3})
with respect to $\lambda _{R}$, yielding
\be
\lambda \; = \; \lambda _{R}(\mu ) \, [ 1 \; + \;
\frac{ \beta _{0} \, \lambda _{R} }{2}
\ln \frac{ \Lambda ^{2} }{ \mu ^{2} } \; + \;
O( \lambda _{R} ^{2} ) ] \; .
\en
But it is this phase of $\phi ^{4}$-theory (where $U(\phi _{c})$ is given
by (\ref{eq:16}) ) that is believed to be trivial~\cite{aiz81}.

However, there is also a second solution of the equations
(\ref{eq:13},\ref{eq:14}). This can most easily be seen from (\ref{eq:13})
for $\phi_{c}=0$; as well as  the solution $M=0$, which corresponds to
the perturbative phase, a second solution
\be
M \; = \; M_{0} \; = \; \mu^{2} \exp \{ \frac{2}{ \beta_{0} \lambda _{R}} \}
\label{eq:17}
\en
exists. The dependence of $M_{0}$ on $\mu $ is exactly the one expected
for a dynamically generated mass in an originally massless
theory~\cite{gro74}, the reason being that $M_{0}$ is renormalisation
group invariant via (\ref{eq:15}).
The effective potential (\ref{eq:14}) for small classical fields
$\phi_{c}$ is
\be
U(\phi _{c}) \; = \;
- \frac{1}{8} \alpha M_{0}^{2} \, + \,  \frac{1}{2} M_{0} \phi _{c}^{2}
\, - \, \frac{1}{4 \alpha }  \phi _{c}^{4} \, - \,
\frac{1}{ 12  \alpha ^{2} M_{0} } \phi_{c}^{6} \, - \,
\frac{1}{ 12 \alpha ^{3} M_{0}^{2} } \phi_{c}^{8} \, + \, \ldots \; ,
\label{eq:18}
\en
where $\alpha = 1/ 16 \pi ^{2} $.
The effective potential obtained from a numerical solution
of (\ref{eq:13}, \ref{eq:14}) is shown in figure 2
for both the perturbative and the non-trivial phase.

Note that the only free parameter in the non-trivial phase
is $M_{0}$ with dimensions $(\hbox{mass})^{2}$. Once it is fixed,
all remaining physical quantities take definite values, and thus
we have a manifestation of dimensional transmutation~\cite{gro74}.
A similar situation occurs in the Gross-Neveu model of self-interacting
fermions where an analogous expansion to (\ref{eq:18}) in terms of the
fermionic condensate is obtained~\cite{la86}.

The first term in (\ref{eq:18}) is the vacuum energy density $\epsilon $
of the non-trivial phase with respect to the perturbative one. It is
negative implying that the perturbative vacuum is unstable.
Moreover, we see from the second terms in this expansion that
a dynamical mass of the scalar particle has been generated.
Note that this mass is large compared to the vacuum energy density.
If we were to assume for example, that $\epsilon $ is of the
electro-weak scale $(\approx ( 200 \, \hbox{GeV})^{4})$, then
this would imply a dynamical  mass of $1192 \, $GeV.
The above considerations show, that the electro-weak energy scale
can nevertheless be consistent with a Higgs mass well above 1 TeV.

\bigskip
\leftline{\bf Acknowledgements: }

We want to thank R.\ Alkofer for useful information.
We are also grateful to R.\ F.\ Langbein for carefully reading this
manuscript and helpful remarks.
\medskip

\vspace{2cm}
{\bf Figure 1: }

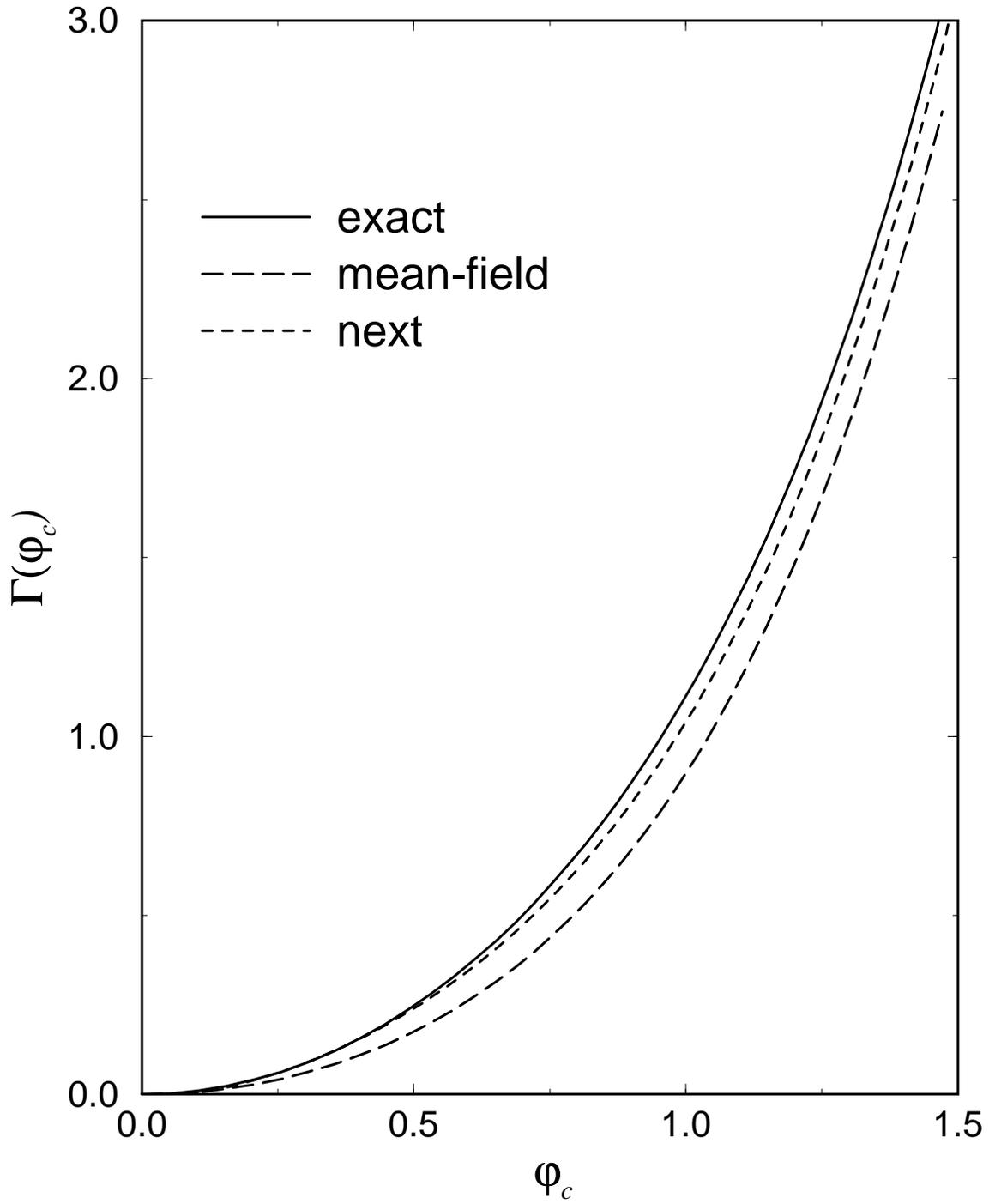
The effective potential in zero dimensions: numerical solution
   (full line), mean-field approximation (long-dashed),
   next to leading order (short dashed).

\vspace{2cm}
{\bf Figure 2: }

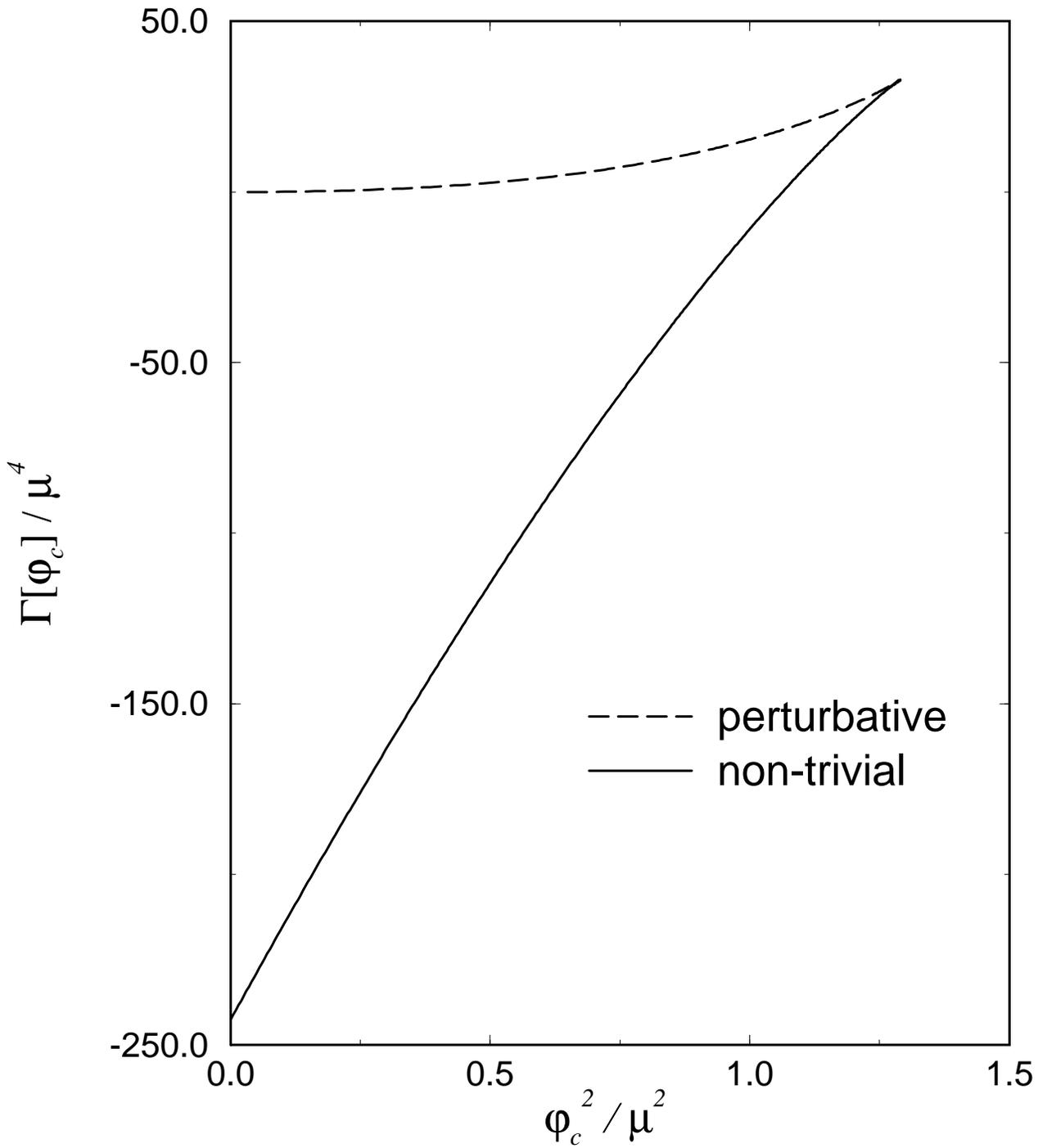
The effective potential in four dimensions: perturbative phase
   (dashed line), non-trivial phase (solid line). (The large $\lambda _{R}$
   has been chosen for aesthetic reasons.)

\end{document}